\documentclass[12pt,a4paper]{article}

\usepackage{latexsym}
\usepackage{setspace}
\usepackage{a4wide}
\usepackage{amsmath}
\usepackage{amsfonts}
\usepackage{amscd}
\usepackage{amssymb}
\usepackage{amsbsy}
\usepackage{mathrsfs}
\usepackage{yfonts}
\usepackage{bbm}
\usepackage[dvips]{epsfig,graphics}
\usepackage{epsfig}
\usepackage{graphicx}
\usepackage{psfrag}
\usepackage{upgreek}
\usepackage{undertilde}
\usepackage{color}
\usepackage{leftidx}
\usepackage{mathbbol}

\DeclareMathAlphabet{\mathpzc}{OT1}{pzc}{m}{it}

\begin{document}

{\color{red}{This document is the Accepted Manuscript version of a Published Work that appeared in final form in International Journal of Solids and Structures, copyright Elsevier after peer review and technical editing by the publisher. To access the final edited and published work see \\ http://www.sciencedirect.com/science/article/pii/S0020768315001626}}

\title{Constitutive modeling of some 2D crystals: graphene, hexagonal BN, MoS$_2$, WSe$_2$ and NbSe$_2$.}

\author{D. Sfyris, G.I. Sfyris,  C. Galiotis}

\maketitle

\begin{abstract}

We lay down a nonlinear elastic constitutive framework for the modeling of some 2D crystals of current interest. The 2D crystals we treat are graphene, hexagonal boron nitride and some metal dichalcogenides: molybdenium disulfide (MoS$_2$), tungsten selenium (WSe$_2$), and niobium diselenide (NbSe$_2$). We first find their arithmetic symmetries by using the theory of monoatomic and diatomic 2-nets. Then, by confinement to weak transformation neighbourhoods and by applying the Cauchy-Born rule we are able to use the symmetries continuum mechanics utilizes: geometric symmetries. We give the complete and irreducible representation for energies depending on an in-plane measure, the curvature tensor and the shift vector. This is done for the symmetry hierarchies that describe how symmetry changes at the continuum level: $\mathcal C_{6 \nu} \rightarrow \mathcal C_{2 \nu} \rightarrow \mathcal C_1$ for monoatomic 2-nets and $\mathcal C_{6 \nu} \rightarrow \mathcal C_{1 \nu} \rightarrow \mathcal C_1$ for diatomic two nets. Having these energies at hand we are able to evaluate stresses and couple stresses for each symmetry regime.

\end{abstract}

\textbf{Keywords:} graphene,  hexagonal BN, MoS$_2$, WSe$_2$, NbSe$_2$, nonlinear elasticity.

\section{Introduction}

Recently, strictly 2D atomic crystals have been isolated from three dimensional layered materials. Novoselov et al (\cite{Novoselovetal2005}) report free standing atomic crystals that can be viewed as individual atomic planes pulled out of the bulk crystal. Using micromechanical cleavage these authors study single layers of graphene, hexagonal boron nitride and some metal dichalcogenides (such as MoS$_2$, NbSe$_2$, and WSe$_2$). 

Motivated by this work, we here lay down a constitutive framework for studying such 2D crystalline materials suitable for the nonlinear elasticity theory. We view graphene as a monoatomic 2-net (\cite{Sfyris-Galiotis2014,Sfyrisetal2014a,Sfyrisetal2014b}), while hexagonal BN, MoS$_2$, NbSe$_2$, and WSe$_2$ are viewed as diatomic 2-nets. The arithmetic symmetry of such 2-nets is well reported in the works of Fadda-Zanzotto (\cite{Fadda-Zanzotto2000,Fadda-Zanzotto2004}) based on the earlier works of Ericksen (\cite{Ericksen1970,Ericksen1979,Ericksen1980,Ericksen1982}), Parry (\cite{Parry1978,Parry1981,Parry1987}), Pitteri (\cite{Pitteri1984,Pitteri1985}) on the definition of arithmetic symmetry. The classification of symmetry in arithmetic classes offers a more stringent classification than crystallogrpahic point groups since conjugacy is taken within the general linear group and not within the orthogonal group (\cite{Pitteri-Zanzotto1996,Fadda-Zanzotto2001b}). 

Monoatomic 2-nets consist of two simple Bravais lattices which have indistinguishable atoms. Graphene belongs to this category (\cite{Sfyris-Galiotis2014,Sfyrisetal2014a,Sfyrisetal2014b}), since it is made of two hexagonal Bravais lattices with carbon atoms occupying atomic positions. On the other hand, diatomic 2-nets consist of two simple Bravais lattices the atomic positions of which are occupied by different atoms. Boron nitride is an example that belongs to this class: one hexagonal Bravais lattice consist of boron atoms only, while the second lattice consists of nitride atoms only. The metal dichalcogenides MoS$_2$, NbSe$_2$, and WSe$_2$ belong to the same category and are also treated in the analysis.  

We here confine the analysis to weak transformation neighbourhoods (\cite{Pitteri-Zanzotto2003}) and use the Cauchy-Born rule (\cite{Ericksen2008}) to lay down the complete and irreducible representation (\cite{Zheng1993,Zheng1994}) for an energy depending on three arguments. The first argument is the surface right Cauchy-Green deformation tensor which is a measure of the in-plane deformations of the 2D crystal. The second argument is the curvature tensor, which introduces out-of-plane deformations and is motivated by the work of Steigmann-Ogden (\cite{Steigmann-Ogden1999}) and earlier approaches on the topic (\cite{Cohen-DeSilva1966,Gurtin-Murdoch1975,Murdoch-Cohen1979}). The third argument that the energy depends on is the shift vector; this is the vector that relates the two lattice. Within these limits arithmetic and geometric symmetries become equivalent which is of particular interest since geometric symmetries are used in continuum mechanics. 

Having the complete and irreducible representation for the energy, we are able to evaluate the surface stress tensor and the surface couple stress tensor. The first being responsible for the in-plane motions, the second for the out-of-plane motions. These measures participate to the momentum and the moment of momentum equations which are the field equations for this problem. From the physical point of view, momentum equation is the force balance for the surface, while moment of momentum equation is the couple balance for the surface. To these field equations one should add the equation ruling the shift vector. Form the physical point of view this equation says that the shift vector adjust so that equilibrium is reached. 

The analysis carries over to cases where symmetry changes for monoatomic 2-nets according to the hierarchies $\mathcal C_{6 \nu} \rightarrow \mathcal C_{2 \nu} \rightarrow \mathcal C_1$. For the diatomic 2-nets symmetry hierarchies change as $\mathcal C_{6 \nu} \rightarrow \mathcal C_{1 \nu} \rightarrow \mathcal C_1$. These groups are the geometric symmetry groups which one derives if the analysis is confined to weak transformation neighbourhoods starting from the arithmetic symmetries. The suitable geometric symmetry group is found by evaluating the eigenvalues for matrices of the arithmetic symmetry and corresponding them to appropriate generators of a geometric symmetry group. We lay down the complete and irreducible representation for the energies for these cases without studying what happens to the transition regime. This is a work in progress in line with similar fundamental works on zirconia (\cite{Truskinovsky-Zanzotto2002}). 

The paper is structured as follows. Section 2 deals with the definition of monoatomic and diatomic 2-nets as well as their symmetries and symmetry hierarchies. Section 3 presents the limitations for the proper transition to the classical continuum viewpoint. Section 4 describes the basic kinematics for a surface energy depending on a surface measure, the curvature tensor and the shift vector. Section 5 desrcibes the way material symmetry should be viewed for the present framework and also gives the field equations. Then, Section 6 gives the complete and irreducible representation for the energy under a specific symmetry group. These symmetry groups are the above mentioned geometric symmetry groups that describe how symmetry breaks for such materials. Surface stress and couple stress tensor can then be evaluated. The article ends up at Section 7 with some concluding remarks and future directions.  

\section{Monoatomic and diatomic 2-nets}   

The importance in the difference between arithmetic and geometric symmetries for crystals stem form the fundamental work of Ericksen (\cite{Ericksen1970}). We refer to the book of Pitteri-Zanzotto (\cite{Pitteri-Zanzotto2003}, and references therein) for a nice exposition of the topic. 

A three dimensional simple lattice in $\mathcal R^3$ is defined as (see e.g. \cite{Pitteri-Zanzotto2003})
\begin{equation}
\mathcal L ({\bf e}_a)=\{ {\bf x} \in \mathcal R^3: {\bf x}=M^a {\bf e}_a, a=1,2,3, M^a \in \mathcal Z \},
\end{equation}
with ${\bf e}_a$ being the lattice vectors and $\mathcal Z$ the space of positive integers. The geometric symmetry group of $\mathcal L$ is (\cite{Pitteri-Zanzotto2003,Ericksen1979})
\begin{eqnarray}
P({\bf e}_a)&=&\{ {\bf Q} \in \mathcal O : \mathcal L({\bf Q} {\bf e}_a)=\mathcal L({\bf e}_a) \nonumber\\
&=&\{ {\bf Q} \in \mathcal O : {\bf Q} {\bf e}_a =m^b_a {\bf e}_b, {\bf m} \in GL(3, \mathcal Z) \},
\end{eqnarray}
where $GL(3, \mathcal Z)$ and $\mathcal O$ are the general linear, and the orthogonal group, respectively. Essentially, this group gives all orthogonal transformations $\bf Q$ that map $\mathcal L$ to itself. Here one finds, for a three dimensional lattice, the 7 crystal systems and the 28 crystallographic point groups continuum mechanics utilizes. 

Due to the fact that conjugacy in $GL(3, \mathcal Z)$ is more stringent than conjugacy in $\mathcal O$, a finer description of symmetry is given by the arithmetic symmetry which is defined as (\cite{Pitteri-Zanzotto2003,Ericksen1979})
\begin{equation}
L({\bf e}_a)=\{ {\bf m} \in GL(3, \mathcal Z): m^b_a {\bf e}_b ={\bf Q} {\bf e}_a, {\bf Q} \in P({\bf e}_a) \}.
\end{equation}
This group gives all distinct types of lattices that are compatible with a given geometric group. Essentially, this is a finer description of symmetry that can differentiate between Bravais lattice types within the same crystal system. 

A multilattice is a generalization of a simple lattice in the sense that it is the finite union of translates of some suitable simple lattice
\begin{equation}
\mathcal M({\bf p}_i, {\bf e}_a)= \cup_{i=0}^{n-1} \mathcal L({\bf p}_i, {\bf e}_a).
\end{equation}
The particular case of a 2-lattice is the union of two simple lattices
\begin{equation}
\mathcal M ({\bf e}_a, {\bf p})=\mathcal L ({\bf e}_a) \cup \{ {\bf p} + \mathcal L({\bf e}_a) \},
\end{equation}
$\mathcal L({\bf e}_a)$ being the simple lattice generated by the basis ${\bf e}_a$, while the vector 
\begin{equation}
{\bf p} =p^1 {\bf e}_1+p^2 {\bf e}_2
\end{equation}
is called the shift vector and gives the separation of the two simple lattices constituting $\mathcal M$. The geometric and arithmetic symmetry groups of multilattices are defined in a similar fashion with the corresponding definitions of simple lattices and we refer to \cite{Pitteri-Zanzotto1998,Fadda-Zanzotto2000,Fadda-Zanzotto2001a,Fadda-Zanzotto2001b,Fadda-Zanzotto2004,Pitteri-Zanzotto2003} for more informations. 

A particular class of a multilattice is the case of monoatomic 2-lattice (net). In this case, two simple lattices constitute the 2-net and atoms at lattice points are indistinguishable, in the sense that they belong to the same species. For the 2 dimensional case the unit cell of monoatomic 2-nets are depicted in Figure 1 (\cite{Fadda-Zanzotto2000}).
\begin{figure}[!htb]
\centering
\includegraphics{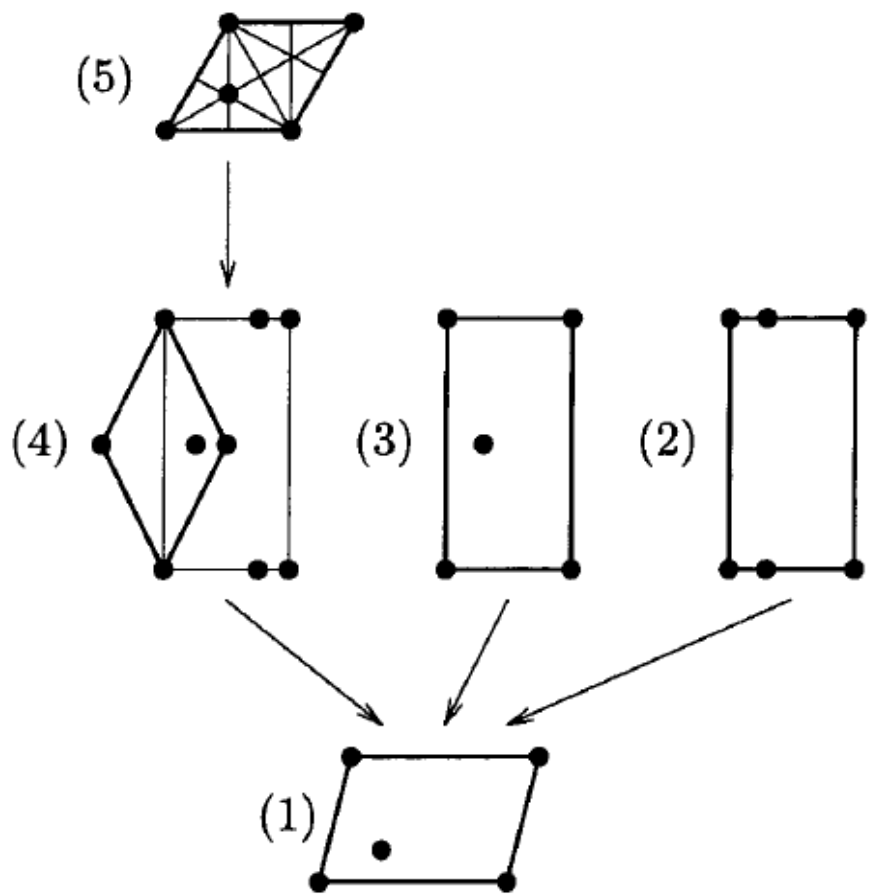}
\caption{Bravais lattices for monoatomic 2-net and their symmetry hierarchies (Figure taken from \cite{Fadda-Zanzotto2000}). Atoms at lattice points of both lattices belong to the same species. }
\label{fig:digraph}
\end{figure}
There are five Bravais lattice for monoatomic 2-nets: (1) oblique, (2) side-rectangular, (3) axis-rectangular, (4) rhombic and (5) hexagonal. 

Graphene is a monoatomic 2-lattice consisting of carbon atoms in all lattice positions that also has a hexagonal unit cell with a shift vector that is depicted in Figure 2. 
\begin{figure}[!htb]
\centering
\includegraphics{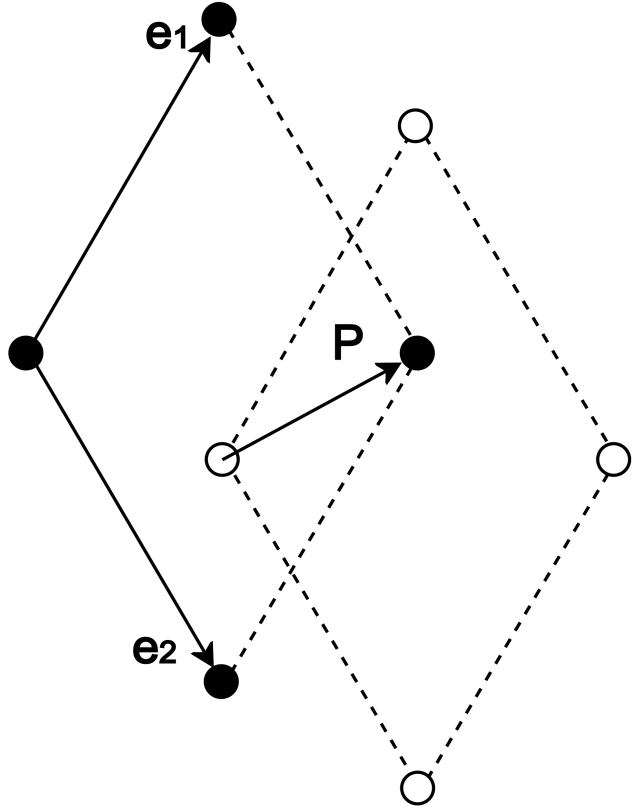}
\caption{A schematic representation of the unit cell of a hexagonal 2-net, depicting the lattice vectors and the shift vector. }
\label{fig:digraph}
\end{figure}
When suitable loading is applied symmetry changes following the scheme (5) $\rightarrow $ (4) $\rightarrow$ (1) that is shown in Figure 1. For the hexagonal case arithmetic symmetries are given by the matrices 
\begin{equation}
\begin{pmatrix}
      -1 & -1 & -1 \\
      1 & 0 & 0 \\
      0 & 0 & 1
\end{pmatrix},
\begin{pmatrix}
      0 & 1 & 0 \\
      1 & 0 & 0 \\
      0 & 0 & 1
\end{pmatrix},
\begin{pmatrix}
      -1 & -1 & -1 \\
      0 & 1 & 0 \\
      0 & 0 & 1
\end{pmatrix},
\end{equation}
\begin{equation}
\begin{pmatrix}
      1 & 0 & 0 \\
      -1 & -1 & -1 \\
      0 & 0 & 1
\end{pmatrix},
\begin{pmatrix}
      1 & 0 & 0 \\
      0 & 1 & 0 \\
      0 & 0 & 1
\end{pmatrix},
\begin{pmatrix}
      0 & 1 & 0 \\
      -1 & -1 & -1 \\
      0 & 0 & 1
\end{pmatrix}.
\end{equation}
To these matrices one should add their counterparts which have -1 at the (3, 3) component of the matrix, since monoatomic 2-nets admit the central inversion as a symmetry operation. The eigenvalues of these matrices are $1, -1, e^{i \frac{\pi}{3}}, e^{-i \frac{\pi}{3}}$. Thus, these matrices describe: the identity transformation, central inversion and rotations by $60, -60$ degrees. The plane group corresponding to this case is p6mm (\cite{Fadda-Zanzotto2000}). 

For the rhombic case the matrices of the arithmetic symmetry read
\begin{equation}
\begin{pmatrix}
      0 & 1 & 0 \\
      1 & 0 & 0 \\
      0 & 0 & 1
\end{pmatrix},
\begin{pmatrix}
      1 & 0 & 0 \\
      0 & 1 & 0 \\
      0 & 0 & 1
\end{pmatrix}.
\end{equation}
The eigenvalues of these matrices are -1 and 1, so they describe central inversion and identity transformation. As above, to these matrices one should add their counterparts which have -1 at the (3, 3) component of the matrix. The plane group for this case is c2mm (\cite{Fadda-Zanzotto2000}).

For the oblique case, the arithmetic symmetry is described by the matrix
\begin{equation}
\begin{pmatrix}
      1 & 0 & 0 \\
      0 & 1 & 0 \\
      0 & 0 & 1
\end{pmatrix},
\end{equation}
together with its counterpart with -1 in the (3, 3) component. The corresponding plane group is p2. 

Diatomic nets are another class of a multilattice: they consist of 2 simple lattices but now the lattice points are occupied by atoms belonging to different species (examples include MoS$_2$, NbSe$_2$ and WSe$_2$ and hexagonal BN). In Figure 2 a diatomic 2-net is depicted with solid and hollow circles referring to different atom species. For the planar case, the unit cell of diatomic 2-nets and their symmetry hierarchies are seen in Figure 3 (\cite{Fadda-Zanzotto2004})  
\begin{figure}[!htb]
\centering
\includegraphics{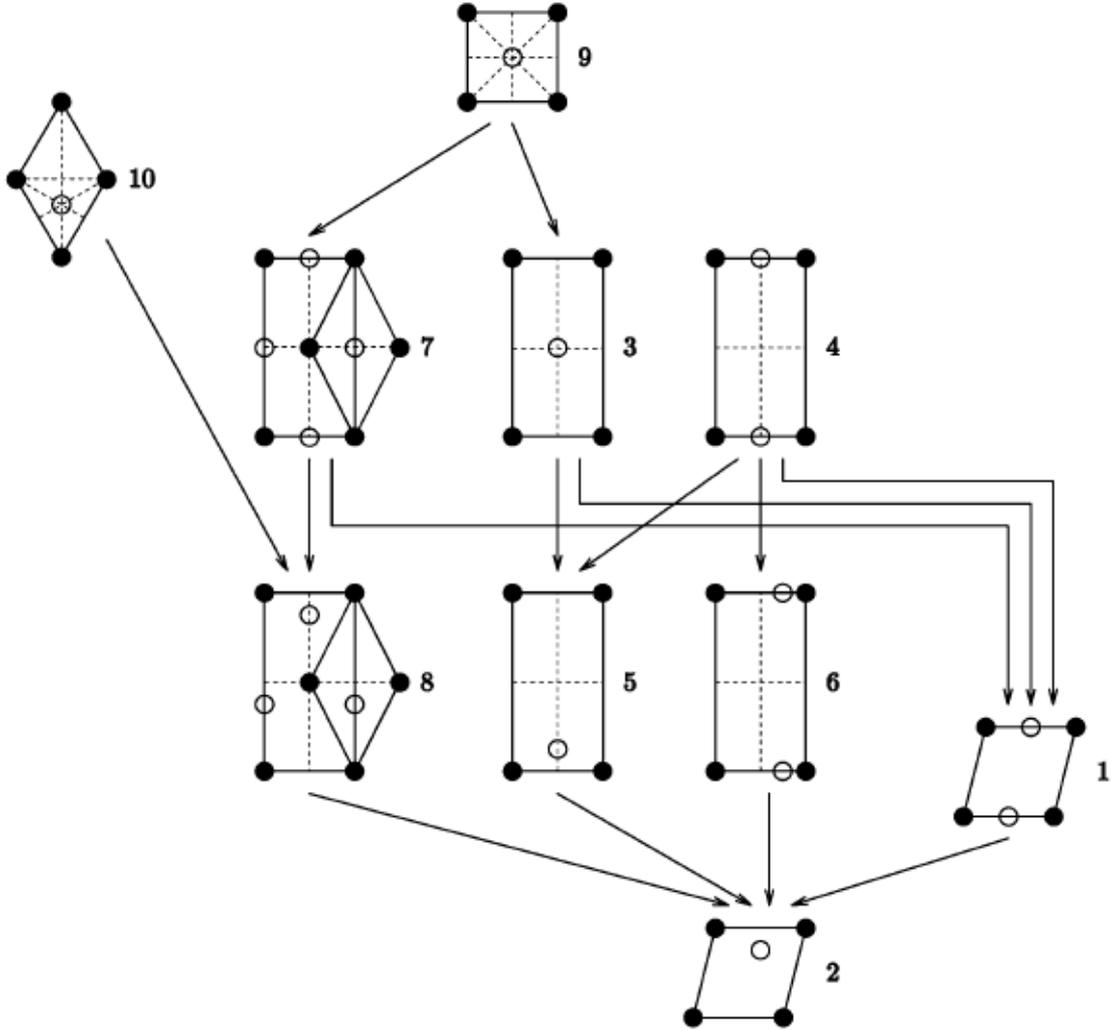}
\caption{Bravais lattices for diatomic 2-nets and their symmetry hierarchies (Figure taken from \cite{Fadda-Zanzotto2004}). Solid circles denote atoms of one species, hollow circles refer to atoms belonging to another species. }
\label{fig:digraph}
\end{figure}
The materials we study here (hexagonal BN, MoS$_2$, WSe$_2$ and NbSe$_2$) are of the hexagonal type (class (10) of Figure 3). Symmetry breaks following the scheme (10) $\rightarrow$ (8) $\rightarrow$ (2) of Figure 3. Diatomic 2-nets differ from their 1-net counterparts since they lack the central inversion. Thus, matrices with -1 at the (3, 3) component of the arithmetic matrices and -1 eigenvalues should be excluded, since they describe central inversion. 
 
\section{Passage to the continuum: the Cauchy-Born rule}

For expressing the discrete nature of a lattice to the continuum scale, we use the Cauchy-Born rule (see e.g. \cite{Ericksen2008} and references therein). Limitations of this rule can be found in \cite{Friesecke-Theil2002} (see also \cite{E-Ming2007}), but in the present work validity of this rule is enforced.

According to this rule, for multilattices, atomic motion of the lattice agrees with the gross deformation, while the shift vector is free to adjust so as to reach equilibrium. We assume the existence of a stored energy function $\phi$ for the multilattice that depends on the current lattice vectors and the shift vector
\begin{equation}
\phi=\phi({\bf e}_a, {\bf p}).
\end{equation} 
The Cauchy-Born rule then states that the reference, ${\bf e}_a^0$, and the current lattice vectors, ${\bf e}_a$, are related according to the formula (\cite{Pitteri-Zanzotto2003})
\begin{equation}
{\bf e}_a={\bf F} {\bf e}_a^0,
\end{equation}
where ${\bf F}=\frac{\partial {\bf x}}{\partial {\bf X}}$ is the well known deformation gradient of continuum mechanics. The shift vector is adjusted so that equilibrium is reached. Using minimization arguments one may show that (\cite{Pitteri-Zanzotto2003})
\begin{equation}
\frac{\partial \phi}{\partial {\bf p}}=0.
\end{equation} 
This equation plays the role of the field equation for the evaluation of the shift vector $\bf p$. Applying the Cauchy-Born rule to the stored energy we write
\begin{equation}
\phi=\phi({\bf F}, {\bf p}). 
\end{equation}

Classical invariance of the energy then would require
\begin{equation}
\phi({\bf F}, {\bf p})=\phi({\bf Q} {\bf F} {\bf H}^T, {\bf p} {\bf H}^T), {\bf H} \in L({\bf e}_a), {\bf Q} \in \mathcal Q, {\bf F} \in \mathcal N_{{\bf e}_a}, 
\end{equation}
but when confined to a weak transformation neighborhood this requirement becomes
\begin{equation}
\phi({\bf F}, {\bf p})=\phi({\bf Q} {\bf F} {\bf H}^T, {\bf p} {\bf H}^T), {\bf H} \in P({\bf e}_a), {\bf Q} \in \mathcal Q, {\bf F} \in \mathcal N_{\bf 1}.
\end{equation}
The action of $\bf H$ is due to material symmetry, while the action of $\bf Q$ is due to frame indifference. Reduction of the arithmetic symmetry group to the geometric can be accomplished when one is confined to weak transformation neighborhoods. In short terms, for simple lattices say that $\mathcal B$ is the 9 dimensional space of all basis ${\bf e}_a$. Then one can prove (\cite{Ericksen1980,Pitteri1984,Pitteri1985}) that there exists a neighborhood $\mathcal N_{\bf 1} \supset \mathcal B$ such that the action of $L({\bf e}_a)$ coincides with that of $P({\bf e}_a)$. Namely, when one is confined to such neighborhoods there is no need to distinguish between geometric and arithmetic symmetry groups. Similar arguments hold for multilattices as well (\cite{Pitteri-Zanzotto2003}). 

So, confined to this neighborhood, material symmetry uses the geometric symmetry group, i.e. the crystal systems continuum mechanics uses. Frame indifference then leads to 
\begin{equation}
\phi=\phi({\bf C}, {\bf p}),
\end{equation}
where ${\bf C}={\bf F}^T {\bf F}$ is the right Cauchy-Green deformation tensor. So, the Cauchy-Born rule allows the transition from a lattice to its continuum counterpart (see Section 6.2, \cite{Pitteri-Zanzotto2003}). What is new for multilattices is the dependence on the shift vector as well (\cite{Pitteri-Zanzotto2003,E-Ming2007}). The motivation for the transformation rule for the shift vector under the action of the symmetry group in eqs. (15, 16) is given in \cite{Sfyris-Galiotis2014}. 

To sum up, there are two crucial assumptions that are necessary for the validity of our model: first is the enforcement of the Cauchy-Born rule and second the confinement of the analysis to weak transformation neighbourhoods. This enables us to work with an energy that has the form  
\begin{equation}
\phi=\phi({\bf C}, {\bf p}),
\end{equation}
augmented properly to take into account bending effects, while symmetry is the one continuum mechanics uses. 

\section{Curvature dependent surface energy}

Arguments of Section 3 pertain to classical three dimensional (bulk) elasticity. In this Section we lay down the kinematics of surface elasticity (\cite{Steigmann-Ogden1999,Wangetal2010}). We assume in the reference configuration a smooth surface A$_0$, described by ${\bf Y}={\bf Y} (\theta^1, \theta^2)$, where $\bf Y$ is the position vector for the point of the surface from the origin and the parameters $\theta^{\alpha}$ are curvilinear coordinates on the surface. Covariant and contravariant vectors are defined by 
\begin{equation}
{\bf  A}_{\alpha} ={\bf Y}_{, \alpha}, \ \  {\bf  A}^{\alpha} \cdot {\bf  A}_{\beta}=\delta^{\alpha}_{\beta},
\end{equation}
where $\delta^{\alpha}_{\beta}$ is the kronecker delta in two dimensions. Deformation of the surface brings point $\bf Y$ to point ${\bf y} (\theta^1, \theta^2)$ on surface A, in the current configuration. The covariant and contravariant vectors related with the surface A are then defined 
\begin{equation}
{\bf  a}_{\alpha} ={\bf y}_{, \alpha}, \ \  {\bf  a}^{\alpha} \cdot {\bf  a}_{\beta}=\delta^{\alpha}_{\beta},
\end{equation}

The linear mapping that maps vectors on the tangent plane of A$_0$ to those of A, is the surface deformation gradient defined by 
\begin{equation}
{\bf F}_s={\bf a}_{\alpha} \otimes {\bf A}^{\alpha}.
\end{equation}
The surface right Cauchy-Green deformation tensor is then defined as 
\begin{equation}
{\bf C}_s={\bf F}_s^T \cdot {\bf F}_s.
\end{equation}
The surface curvature tensors in the reference and the current configuration can be expressed as 
\begin{eqnarray}
&& {\bf b}_0=b_{0 \alpha \beta}  {\bf  A}^{\alpha} \otimes  {\bf  A}^{\beta}, \\
&& {\bf b}=b_{\alpha \beta}  {\bf  a}^{\alpha} \otimes  {\bf  a}^{\beta}.
\end{eqnarray}
Surface divergence is defined by 
\begin{equation}
\nabla_s ()=\nabla()-{\bf n} ({\bf n} \cdot \nabla ()),
\end{equation}
where $\nabla$ is the common divergence operator in three dimensions, while $\bf n$  and $\bf N$are the outward unit normals on surface A and A$_0$, respectively. 

A curvature dependent surface energy is an energy of the form (\cite{Steigmann-Ogden1999,Cohen-DeSilva1966,Murdoch-Cohen1979})
\begin{equation}
W=W({\bf C}_s, {\bf b}_0). 
\end{equation}
So, aside from the in-surface strain measure ${\bf C}_s$, dependence on ${\bf b}_0$ is assumed. This dependence allows the modeling of bending effects since it takes into account out-of-plane deformations. For a monolayer 2D crystal this energy should be augmented with the dependence on the shift vector which gives
\begin{equation}
W=W({\bf C}_s, {\bf b}_0, {\bf p}),
\end{equation}
where $\bf p$ is the shift vector. The need for the dependence on the shift vector stems from the fact that monoatomic and diatomic 2-nets are multilattices.

\section{Material symmetry and field equations}

Material symmetry for curvature dependent surface energies is a subject tackled elegantly by Steigmann and Ogden (\cite{Steigmann-Ogden1999} Section 6) based on the earlier work of Murdoch and Cohen (\cite{Murdoch-Cohen1979}). For such energies, elements of the symmetry group are pairs which leave the response of the surface invariant to superposed deformations. Steigmann and Ogden (\cite{Steigmann-Ogden1999}) concluded that for surfaces with constant non-negative curvature (namely, for planes and spheres) amenability to available representation theory is possible when the symmetry group reads $\{ {\bf R}, {\bf 0} \}$ while for the energy it then holds
\begin{equation}
W({\bf C}_s, {\bf b}_0)=W({\bf R} {\bf C}_s {\bf R}^T, {\bf R} {\bf b}_0 {\bf R}^T), {\bf R} {\bf R}^T = {\bf I}, \ \ \textrm{det} {\bf R}=1.
\end{equation}

The form of the energy should be augmented by taking into account dependence on the shift vector. So, collectively, the action of the symmetry group for a curvature dependent surface energy reads
\begin{equation}
W({\bf C}_s, {\bf b}_0, {\bf p})=W({\bf R} {\bf C}_s {\bf R}^T, {\bf R} {\bf b}_0 {\bf R}^T, {\bf p} {\bf R}^T). 
\end{equation}
The motivation for the transformation rule of the shift vector is given in \cite{Sfyris-Galiotis2014}.

Field equations for materials described by curvature dependent surface energies are studied in \cite{Chhapadiaetal2011}. According to these authors, the momentum equation for the static case and in the absence of body forces reads
\begin{equation}
\boldsymbol \sigma^{\textrm{bulk}} \cdot {\bf n} + \nabla_S {\bf T}_S=0,
\end{equation}
where ${\bf T}_S$ is the surface first Piola-Kirchhoff stress tensor defined by (\cite{Steigmann-Ogden1999})
\begin{equation}
{\bf T}_S=\frac{\partial \bar{W}}{\partial {\bf F}_S}
\end{equation}
when $W=\bar{W}({\bf F}_S, {\bf b}_0, {\bf p})$, while $\boldsymbol \sigma^{\textrm{bulk}}$ is the Cauchy stress tensor for the bulk material surrounding the surface. Here since we speak about free standing surfaces, there is no bulk material, so the bulk stress tensor should be set equal to zero, $\boldsymbol \sigma^{\textrm{bulk}}={\bf 0}$. The surface first Piola-Kirchhoff stress tensor is related to the second Piola-Kirchhoff surface stress tensor, ${\bf S}_S$, according to the formula
\begin{equation}
{\bf S}_S={\bf F}_S^{-1} \cdot  {\bf T}_S.
\end{equation} 
The second Piola-Kirchhoff stress tensor can also be written as
\begin{equation}
{\bf S}_S=\frac{\partial W}{\partial {\bf C}_S}.
\end{equation}
The momentum equation in the absence of body forces and inertia can be expressed using the second Piola-Kirchhoff stress tensor as
\begin{equation}
\bar{\nabla}_S {\bf S}_S={\bf 0},
\end{equation}
where $\bar{\nabla}_S ()= \nabla^{\bf C}_S-{\bf N} ({\bf N} \cdot \nabla^{\bf C}())  $ is the surface gradient when the bulk gradient $\nabla^{\bf C}$ is taken with respect to the metric $\bf C$ of the bulk. In this respect, the bulk divergence in eq. (25) is taken with respect to the metric $\bf G$ of the bulk reference configuration. Essentially, eq. (34) is the surface analog of the momentum equation written using the second Piola-Kirchhoff stress tensor in classical elasticity (\cite{Marsden-Hughes,Simo-Marsden1984,Simoetal1988}). From the physical point of view, the momentum equation is the force balance for the surface.

The moment of momentum balance, in the absence of body couples and inertia reads (\cite{Chhapadiaetal2011}) when setting $\boldsymbol \sigma^{\textrm{bulk}}={\bf 0}$
\begin{equation}
\nabla_S {\bf m}_S-\nabla_S ({\bf F}_S^{-1} \cdot  {\bf S}_S \times {\bf y})={\bf 0}, 
\end{equation}
where the surface couple stress tensor is defined as (\cite{Steigmann-Ogden1999})
\begin{equation}
{\bf m}_S=\frac{\partial W}{\partial {\bf C}_S}. 
\end{equation}
The symbol $\times$ in eq. (35) denotes the cross product of the three dimensional space. The moment of momentum equation renders the couple balance for the surface. 

For the shift vector the field equation reads (\cite{Pitteri-Zanzotto2003,E-Ming2007})
\begin{equation}
\frac{\partial W}{\partial {\bf p}}={\bf 0}.
\end{equation}
Form the physical point of view, this equation says that the shift vector adjusts in sucha way that equilibrium is reached. 

\section{Constitutive modeling for each symmetry regime}

To obtain the appropriate geometric symmetry group we first evaluate eigenvalues for the matrices of the arithmetic symmetry group. We then correspond these eigenvalues to appropriate generators of the geometric symmetry group. This is done for the hierarchies (5)$\rightarrow$(4)$\rightarrow$(1) for monoatomic 2-nets and (10)$\rightarrow$(8)$\rightarrow$(1) for diatomic 2-nets. 

\subsection{Hexagonal lattice}

For hexagonal lattices the arithmetic symmetry group is given by the matrices of eq. (7, 8) for monoatomic 2-nets. Evaluation of the eigenvalues gives $1, -1, e^{i \frac{\pi}{3}}, e^{-i \frac{\pi}{3}}$ which describe identity transformation, central inversion and rotation by -60, 60 degrees. The space group is p6mm. For their diatomic counterparts central inversion is excluded. The corresponding crystallographic point group has generators ${\bf R} (\frac{2 \pi}{6})$ and ${\bf R}_{\bf j}$ and belongs to class 10 according to the classification of Zheng (\cite{Zheng1993,Zheng1994}) and is denoted by $\mathcal C_{6 \nu}$. ${\bf R} (\theta)$ denotes a rotation of angle $\theta$ and ${\bf R}_{\bf j}$ is a two-dimensional reflection transformation. Diatomic 2-nets have the same geometric symmetry group, $\mathcal C_{6 \nu}$, since this group does not admit central inversion. 

The structure tensor for $\mathcal C_{6 \nu}$ is denoted by ${\bf P}_6$ and defined by (\cite{Zheng1993})
\begin{equation}
{\bf P}_6=Re({\bf a}_1+i {\bf a}_2)^6,
\end{equation}
or equivalently as 
\begin{equation}
{\bf P}_6={\bf M} \otimes {\bf M} \otimes {\bf M}-({\bf N} \otimes {\bf M}  \otimes {\bf N}+{\bf N} \otimes {\bf N} \otimes {\bf M}),
\end{equation}
where ${\bf M}={\bf a}_1 \otimes {\bf a}_1- {\bf a}_2 \otimes {\bf a}_2$, ${\bf N}={\bf a}_1 \otimes {\bf a}_2 - {\bf a}_2 \otimes {\bf a}_1$, ${{\bf a}_1, {\bf a}_2}$, $\{ {\bf a}_1, {\bf a}_2  \}$ an orthonormal basis vector. It can also be written as  
\begin{equation}
{\bf P}_6=Re[e^{i 6 \theta} ({\bf c}_1+i {\bf c}_2)^6],
\end{equation} 
where ${\bf c}_1=cos(\theta){\bf a}_1+sin(\theta) {\bf a}_2$, ${\bf c}_2=-sin(\theta){\bf a}_1+cos(\theta) {\bf a}_2$. This tensor is an irreducible tensor since $\mathcal C_{6 \nu}$ is compact. In two dimensions it has only two independent components (\cite{Zheng1994})
\begin{equation}
{\bf P}_{111111}=\textrm{cos}(6 \theta), \ \ {\bf P}_{211111}=\textrm{sin}(6 \theta).
\end{equation}
These two components introduce the anisotropy and they model the zig-zag and armchair direction. Since $\theta=\frac{2 \pi}{6}$ we have $P_{111111}=cos(2 \pi)=1, P_{211111}=sin(2 \pi)=0 $.

Thus, for this case we have using curvature dependent surface energy 
\begin{equation}
W=W_{anisotropic}({\bf C}_S, {\bf b}_0, {\bf p}),
\end{equation}
where the anisotropy stems from the fact that the symmetry group is not the full orthogonal group. Reduction to an isotropic function is done through the use of the principle of isotropy of space (\cite{Zheng1994}), that gives
\begin{equation}
W=W_{anisotropic}({\bf C}_S, {\bf b}_0, {\bf p})=W_{isotropic}({\bf C}_S, {\bf b}_0, {\bf P}_6, {\bf p}). 
\end{equation}
Thus, we take an isotropic function at the expense of using the structure tensor (that describes the anisotropy) as an addittional argument. 

The complete and irreducible representation of such a scalar function under the group $\mathcal C_{6 \nu}$ consists of the following quantities (\cite{Zheng1994,Zheng1993})
\begin{eqnarray}
&&I_1=\textrm{tr}{\bf C}_S, \ I_2=\textrm{det}{\bf C}_S, \ I_3=\textrm{tr}(\Pi^{{\bf C}_S}_6 {\bf C}_S ), \ I_4=\textrm{tr}({\bf C}_S {\bf b}_0), \nonumber\\
&&I_5=\textrm{tr}(\Pi^{{\bf b}_0}_6 {\bf b}_0), \  I_6=\textrm{tr}{\bf b}_0, \ I_7=\textrm{det}{\bf b}_0, \ I_8={\bf p} \cdot {\bf C}_S {\bf p}, \nonumber\\
&&I_9={\bf p} \cdot {\bf b}_0 {\bf p}, \  I_{10}={\bf p} \cdot \pi^{{\bf p}}_6, \ I_{11}={\bf p} \cdot {\bf p}, \ I_{12}=\textrm{tr}(\Pi^{\bf p}_6 {\bf C}_S), \ I_{13}=\textrm{tr}(\Pi^{\bf p}_6 {\bf b}_0).
\end{eqnarray}
Thus, in general for such a model we have the following expression for the energy 
\begin{equation}
W=\tilde{W}(I_i), \ i=1,2,...,13. 
\end{equation}
The term $\Pi^{\bf A}_6$ for a symmetric tensor of second order $\bf A$ is defined as (\cite{Zheng1994}, indices ranging from 1 to 2)
\begin{equation}
\Pi^{\bf A}_6=P_{ijklmn} A_{kl} A_{mn} {\bf c}_i \otimes {\bf c}_j,
\end{equation}
and renders a second order tensor. The term $\pi^x_6$ with respect to the vector $\bf z$ is defined as 
\begin{equation}
\pi^{\bf z}_6=P_{ijklmn} z_j z_k z_l z_m z_n {\bf c}_i,
\end{equation}
while for $\Pi^{\bf z}_6$ we have
\begin{equation}
\Pi^{\bf z}_6=P_{ijklmn} z_k z_l z_m z_n {\bf c}_i \otimes {\bf c}_j.
\end{equation}

The material parameters related to $I_6, I_7$ describe pure bending effects since det${\bf b}_0$, tr${\bf b}_0$ are the mean and the Gaussian curvature of the surface, respectively. The term related with $I_5$ describes the effect of the armchair and the zig-zag direction at bending. The parameters related with $I_1, I_2$ are related to pure stretching, while those related with the term $I_3$ describe the anisotropy (zig-zag, armchair directions) to stretching. The parameter related to $I_4$ describes coupling between bending and stretching response. The terms $I_8, I_9$ describe the effect of the in plane and the out of plane deformations, respectively, on the shift vector. The term $I_{10}$ describes the way anisotropy affects the shift vector, while $I_{11}$ describe changes related with the shift vector solely. Terms $I_{12}, I_{13}$ are related to coupling of anisotropy with the shift vector for the in plane and the out of plane deformations, respectively. 

The surface stress tensor and the surface couple stress tensor are evaluated by determining the derivatives
\begin{equation}
{\bf S}_S=\frac{\partial \tilde{W}}{\partial {\bf C}_S}, \ \ {\bf m}_S=\frac{\partial \tilde{W}}{\partial {\bf b}_0}.
\end{equation}
Also, the field equation for the shift vector is $\frac{\partial \tilde{W}}{\partial {\bf p}}$ as shown in eq. (37). Using the expressions of eq. (44) in eq. (49), after some calculations we obtain 
\begin{eqnarray}
{\bf S}_S&=&\frac{\partial \tilde{W}}{\partial I_1} {\bf I} + \frac{\partial \tilde{W}}{\partial I_2} [ \textrm{tr}({\bf C}_S) {\bf 1} -{\bf C}_S ]+3 \frac{\partial \tilde{W}}{\partial I_3} {\bf P}_6 : ({\bf C}_S \otimes  {\bf C}_S)+ \frac{\partial \tilde{W}}{\partial I_4} {\bf b}_0 \nonumber\\ 
&&+\frac{\partial \tilde{W}}{\partial I_8} {\bf p} \otimes {\bf p}+\frac{\partial \tilde{W}}{\partial I_{12}}\Pi^{\bf p}_6, \\
{\bf m}_S&=&\frac{\partial \tilde{W}}{\partial I_6} {\bf I} + \frac{\partial \tilde{W}}{\partial I_7} [ \textrm{tr}({\bf b}_0) {\bf 1} -{\bf b}_0 ]+3 \frac{\partial \tilde{W}}{\partial I_5} {\bf P}_6 : ({\bf b}_0 \otimes  {\bf b}_0)+ \frac{\partial \tilde{W}}{\partial I_4} {\bf C}_S \nonumber\\
&& +\frac{\partial \tilde{W}}{\partial I_9} {\bf p} \otimes {\bf p}+\frac{\partial \tilde{W}}{\partial I_{13}} \Pi^{\bf p}_6, \\
\frac{\partial W}{\partial {\bf p}}&=&2\frac{\partial \tilde{W}}{\partial I_8} {\bf C}_S {\bf p}+2\frac{\partial \tilde{W}}{\partial I_9} {\bf b}_0 {\bf p}+6 \frac{\partial \tilde{W}}{\partial I_{10}} {\bf P}_6 \bullet ({\bf p} \otimes {\bf p} \otimes {\bf p} \otimes {\bf p} \otimes {\bf p}) + \frac{\partial \tilde{W}}{\partial I_{11}} {\bf p} \nonumber\\
&&+4 \frac{\partial \tilde{W}}{\partial I_{12}} [{\bf P}_6 : ( {\bf p} \otimes {\bf p} \otimes {\bf p})]: {\bf C}_S+4 \frac{\partial \tilde{W}}{\partial I_{12}} [{\bf P}_6 : ( {\bf p} \otimes {\bf p} \otimes {\bf p})]: {\bf b}_0.
\end{eqnarray}
By making the simplest possible assumption that $\tilde{W}$ is linear with respect to the invariants $I_i, i=1,2,3,...,13$ we take 
\begin{eqnarray}
{\bf S}_S&=&\alpha{\bf I} + \beta [ \textrm{tr}({\bf C}_S) {\bf 1} -{\bf C}_S ]+3 \gamma {\bf P}_6 : ({\bf C}_S \otimes  {\bf C}_S)+ \delta {\bf b}_0+\theta  {\bf p} \otimes {\bf p}+\rho \Pi^p_6, \\
{\bf m}_S&=&\epsilon {\bf I} + \zeta [ \textrm{tr}({\bf b}_0) {\bf 1} -{\bf b}_0 ]+3 \eta {\bf P}_6 : ({\bf b}_0 \otimes  {\bf b}_0)+ \delta {\bf C}_S+\iota {\bf p} \otimes {\bf p}+\tau \Pi^p_6, \\
\frac{\partial W}{\partial {\bf p}}&=&\theta {\bf C}_S {\bf p}+\iota {\bf b}_0 {\bf p}+6 \lambda {\bf P}_6 \bullet( {\bf p} \otimes {\bf p} \otimes {\bf p} \otimes {\bf p} \otimes {\bf p})+\xi {\bf p} \nonumber\\
&& +4 \rho  [{\bf P}_6 : ( {\bf p} \otimes {\bf p} \otimes {\bf p})]: {\bf C}_S +4 \tau [{\bf P}_6 : ( {\bf p} \otimes {\bf p} \otimes {\bf p})]: {\bf b}_0.
\end{eqnarray}
The Greek letters $\alpha, \beta, \gamma, \delta, \theta, \rho, \epsilon, \zeta, \eta, \iota, \tau, \lambda, \xi$ are material parameters to be determined by experiments. 

Written with respect to indices ranging from 1 to 2 these formulas are
\begin{eqnarray}
S_{S_{AB}}&=&\alpha \delta_{AB}+\beta [tr({\bf C}_S) \delta_{AB}-C_{S_{AB}}]+3 \gamma P_{ABCDEF} C_{S_{EF}} C_{S_{CD}} +\delta d_{0_{AB}} \nonumber\\
&& +\theta p_A p_B +\rho P_{ABCDEF} p_C p_D p_E p_F, \\
m_{S_{AB}}&=&\epsilon \delta_{AB}+\zeta [tr({\bf b}_0) \delta_{AB}-b_{0_{AB}}]+3 \eta P_{ABCDEF} b_{0_{EF}} b_{0_{CD}} +\delta C_{S_{AB}} \nonumber\\
&& +\iota p_A p_B +\tau P_{ABCDEF} p_C p_D p_E p_F, \\
\frac{\partial W}{\partial p_A}&=& \theta C_{S_{AB}} p_A +\iota b_{0_{AB}} p_B+6 \lambda P_{ABCDEF} p_B p_C p_D p_E p_F+\xi p_A \nonumber\\
&&+4 \rho P_{ABCDEF} p_D p_E p_F C_{S_{BC}}+4 \tau P_{ABCDEF} p_D p_E p_F b_{0_{BC}}. 
\end{eqnarray}

The elasticities of this model are given by the following fourth order tensors
\begin{equation}
\mathcal A=\frac{\partial^2 W}{\partial {\bf C}_S^2}, \ \ \mathcal B=\frac{\partial^2 W}{\partial {\bf b}_0^2}, \ \ \mathcal C=\frac{\partial^2 W}{\partial {\bf C}_S \partial {\bf b}_0}.
\end{equation}
Quantities of the first term are related to the in-plane motion, the second to the out-of-plane while the third to the coupling between in-plane and out-of-plane motions. All in all, modeling of the hexagonal lattice at the continuum level for monoatomic and diatomic 2-nets requires specification of 13 material parameters; these are the Greek letters $\alpha, \beta, \gamma, \delta, \theta, \rho, \epsilon, \zeta, \eta, \iota, \tau, \lambda, \xi$. 

\subsection{First braking of symmetry}

Breaking of symmetry is dictated by Figures 1 and 3 which describe the symmetry hierarchies for monoatomic and diatomic 2-nets. We again follow the approach of evaluating the eigenvalues of the matrices of the arithemtic symmetry group and finding the corresponding geometric symmetry groups with generators describing the same symmetry operation. For the first braking of symmetry one has to distinguish between monoatomic and diatomic 2-nets.

\subsubsection{Monoatomic 2-nets}

For monoatomic 2-nets the rhombic unit cell has arithmetic symmetry group described by eq. (9), and the corresponding eigenvalues are 1 and -1. The corresponding geometric symmetry group is the group $\mathcal C_{2 \nu}$ (no. 4 in the classification adopted in \cite{Zheng1993,Zheng1994}) with generators ${\bf R} (\pi), {\bf R}_{\bf j}$. This case corresponds to orthotropy in two dimensions and the structure tensor in this case is the tensor ${\bf M}={\bf a}_1 \otimes {\bf a}_1 -{\bf a}_2 \otimes {\bf a}_2$. 

Thus, in this case we have
\begin{equation}
W_{anisotropic}({\bf C}_S, {\bf b}_0, {\bf p})=W_{isotropic}({\bf C}_S, {\bf b}_0, {\bf p}, {\bf M}).
\end{equation}
The complete and irreducible representation of such a scalar function under the group $C_{2 \nu}$ consists of the following quantities (\cite{Zheng1993,Zheng1994})
\begin{eqnarray}
&&I_1=\textrm{tr}{\bf C}_S, \ I_2=\textrm{det}{\bf C}_S, \ I_3=\textrm{tr}({\bf M} {\bf C}_S ), \ I_4=\textrm{tr}({\bf C}_S {\bf b}_0), \nonumber\\
&&I_5=\textrm{tr}({\bf M} {\bf b}_0), \  I_6=\textrm{tr}{\bf b}_0, \ I_7=\textrm{det}{\bf b}_0, \ I_8={\bf p} \cdot {\bf p}, \nonumber\\
&&I_9={\bf p} \cdot {\bf M} {\bf p}, \ I_{10}={\bf p} \cdot {\bf C}_S {\bf p}, \ I_{11}={\bf p} \cdot {\bf b}_0 {\bf p}.
\end{eqnarray}
Thus, for the energy we obtain
\begin{equation}
W=\tilde{W}(I_i), \ i=1,2,...,11. 
\end{equation}

The surface stress and surface couple stress that correspond to this energy are calculated as
\begin{eqnarray}
&&{\bf S}_S=\frac{\partial \tilde{W}}{\partial I_1} {\bf I} + \frac{\partial \tilde{W}}{\partial I_2} [ \textrm{tr}({\bf C}_S) {\bf 1} -{\bf C}_S ]+ \frac{\partial \tilde{W}}{\partial I_3} {\bf M} : ({\bf C}_S \otimes  {\bf C}_S)+ \frac{\partial \tilde{W}}{\partial I_4} {\bf b}_0+\frac{\partial \tilde{W}}{\partial I_{10}} {\bf p} \otimes {\bf p}  . \\
&&{\bf m}_S=\frac{\partial \tilde{W}}{\partial I_4} {\bf I} + \frac{\partial \tilde{W}}{\partial I_7} [ \textrm{tr}({\bf b}_0) {\bf 1} -{\bf b}_0 ]+ \frac{\partial \tilde{W}}{\partial I_5} {\bf M} : ({\bf b}_0 \otimes  {\bf b}_0)+ \frac{\partial \tilde{W}}{\partial I_4} {\bf C}_S + \frac{\partial \tilde{W}}{\partial I_{11}} {\bf p} \otimes {\bf p}.
\end{eqnarray}
For the term $\frac{\partial W}{\partial {\bf p}}$ we have
\begin{equation}
\frac{\partial W}{\partial {\bf p}}=\frac{\partial \tilde{W}}{\partial I_8} {\bf p} +\frac{\partial \tilde{W}}{\partial I_9} {\bf M}  {\bf p}+\frac{\partial \tilde{W}}{\partial I_{10}} {\bf C}_S   {\bf p}+\frac{\partial \tilde{W}}{\partial I_{11}} {\bf b}_0  {\bf p}.
\end{equation}
By making the simplest possible assumption that $\tilde{W}$ is linear with respect to the invariants $I_i, i=1,2,3,..., 11$ we take 
\begin{eqnarray}
&&{\bf S}_S=\alpha{\bf G} + \beta [ \textrm{tr}({\bf C}_S) {\bf 1} -{\bf C}_S ]+ \gamma {\bf M} : ({\bf C}_S \otimes  {\bf C}_S)+ \delta {\bf b}_0+\theta {\bf p} \otimes {\bf p}, \\
&&{\bf m}_S=\epsilon {\bf G} + \zeta [ \textrm{tr}({\bf b}_0) {\bf 1} -{\bf H}_S ]+ \eta {\bf M} : ({\bf b}_0 \otimes  {\bf b}_0)+ \delta {\bf C}_S+\iota {\bf p} \otimes {\bf p}
\end{eqnarray}
and 
\begin{equation}
\frac{\partial W}{\partial {\bf p}}=\lambda {\bf p} +\mu {\bf M}  {\bf p}+\theta {\bf C}_S  {\bf p}+\iota {\bf b}_0  {\bf p}.
\end{equation}
Now the effect of anisotropy at the level of the constitutive law is introduced throught the terms where the structure tensor, $\bf M$, is present. Namely the material parameters $\gamma, \eta, \mu$ measure the effect of anisotropy at the in-plane motion, the out-of-plane motion and the shift vector, respectively. For monoatomic 2-nets the number of independent material constants to be observed and measured in experiments are 11: the Greek letters $\alpha, \beta, \gamma, \delta, \theta, \epsilon, \zeta, \eta, \iota, \lambda, \mu$.

\subsubsection{Diatomic 2-nets}

Diatomic 2-nets lack central inversion. Thus, we assume that the corresponding geometric symmetry group is the group $\mathcal C_{1 \nu}$ (no.3 in the classification adopted in \cite{Zheng1993,Zheng1994}) with generators ${\bf R}_{\bf j}$. In this case the structure tensor is the vector ${\bf a}_1$. So, for this case energy reads
\begin{equation}
W_{anisotropic}({\bf C}_S, {\bf b}_0, {\bf p})=W_{isotropic}({\bf C}_S, {\bf b}_0, {\bf p}, {\bf a}_1).
\end{equation}
The complete and irreducible representation of such a scalar function under the group $C_{1 \nu}$ consists of the following quantities (\cite{Zheng1993})
\begin{eqnarray}
&&I_1=\textrm{tr} {\bf C}_S, \ I_2=\textrm{tr} ({\bf M} {\bf C}_S), \ I_3=\textrm{tr} {\bf C}_S^2, \ I_4={\bf p} \cdot {\bf a}_1, \ I_5={\bf p} \cdot {\bf p} \nonumber\\
&&I_6=\textrm{tr} ({\bf C}_S {\bf b}_0), \ I_7= {\bf p} \cdot {\bf C}_S {\bf a}_1, \ I_8=\textrm{tr} {\bf b}_0, \ I_9=\textrm{tr}({\bf M} {\bf b}_0), \nonumber\\
&&I_{10}= \textrm{tr} {\bf b}_0^2, \ I_{11}={\bf p} \cdot {\bf b}_0 {\bf a}_1. 
\end{eqnarray}
Thus, for the energy we obtain
\begin{equation}
W=\tilde{W}(I_i), \ i=1,2,...,11. 
\end{equation}

The surface stress tensor for this case then reads
\begin{eqnarray}
{\bf S}=\frac{\partial W}{\partial {\bf C}_S}&&=\frac{\partial \tilde{W}}{\partial I_1} {\bf I}+\frac{\partial \tilde{W}}{\partial I_2} {\bf M}+\frac{\partial \tilde{W}}{\partial I_3} [(\textrm{tr} {\bf C}_S ){\bf I}-{\bf C}_S ]+\frac{\partial \tilde{W}}{\partial I_6} {\bf b}_0 +\frac{\partial \tilde{W}}{\partial I_7} {\bf p} \otimes {\bf a}_1 \nonumber\\
&&=\alpha {\bf I}+\beta {\bf M}+\gamma [(\textrm{tr} {\bf C}_S ){\bf I}-{\bf C}_S ]+\delta {\bf b}_0 +\epsilon {\bf p} \otimes {\bf a}_1,
\end{eqnarray}
and the energy is assumed to be linear with respect to the invariants. For the surface couple stress under the same energy assumption we then have
\begin{eqnarray}
{\bf m}=\frac{\partial W}{\partial {\bf b}_0}&&=\frac{\partial \tilde{W}}{\partial I_6} {\bf C}_S + \frac{\partial \tilde{W}}{\partial I_8} {\bf I}+\frac{\partial \tilde{W}}{\partial I_9} {\bf M}+\frac{\partial \tilde{W}}{\partial I_{10}}  [(\textrm{tr} {\bf b}_0 ){\bf I}-{\bf b}_0 ]+ \frac{\partial \tilde{W}}{\partial I_{11}} {\bf p} \otimes {\bf a}_1 \nonumber\\
&&=\delta {\bf C}_S + \zeta {\bf I}+\eta {\bf M}+\theta  [(\textrm{tr} {\bf b}_0 ){\bf I}-{\bf b}_0 ]+ \iota {\bf p} \otimes {\bf a}_1. 
\end{eqnarray} 
For the term $\frac{\partial W}{\partial {\bf p}}$ we find
\begin{eqnarray}
\frac{\partial W}{\partial {\bf p}}&&=\frac{\partial \tilde{W}}{\partial I_4} {\bf a}_1+\frac{\partial \tilde{W}}{\partial I_5} {\bf p}+\frac{\partial \tilde{W}}{\partial I_7} ({\bf C}_S \cdot {\bf a}_1)+\frac{\partial \tilde{W}}{\partial I_{11}} ({\bf b}_0 \cdot {\bf a}_1) \nonumber\\
&&=\kappa  {\bf a}_1+\lambda {\bf p}+\epsilon ({\bf C}_S \cdot {\bf a}_1)+\iota ({\bf b}_0 \cdot {\bf a}_1)
\end{eqnarray}

The material parameters related with $I_1, I_3$ describe pure stretching, while those related with $I_2$ describe the effect of anisotropy to pure stretching. Pure bending characteristics are introduced throught terms $I_8, I_{10}$, while the effect of anisotropy to pure bending is described by $I_9$. Combined stretching and bending effects are given by $I_6$, while the effect of anisotropy to the shift vector is throught the term $I_4$. Terms $I_7, I_{11}$ describe how anisotropy affect the combined response of the shift vector with stretching and bending, respectively. Finally, $I_5$ is a term for the shift vector solely.  For diatomic 2-nets the number of independent material constants to be observed and measured in experiments are 11: the Greek letters $\alpha, \beta, \gamma, \delta, \theta, \epsilon, \zeta, \eta, \iota, \kappa, \lambda$.

\subsection{Second breaking of symmetry}

In this case both monoatomic and diatomic 2-nets have as geometric symmetry group the group $\mathcal C_1$ (no. 1 in the classifiaction adopted by \cite{Zheng1993,Zheng1994}). The structure tensors for this group are the vectors ${\bf a}_1, {\bf a}_2$. So, for the energy we have 
\begin{equation}
W_{anisotropic}({\bf C}_S, {\bf b}_0, {\bf p})=W_{isotropic}({\bf C}_S, {\bf b}_0, {\bf p}, {\bf a}_1, {\bf a}_2).
\end{equation}
Invariants for this energy are then
\begin{eqnarray}
&&I_1=\textrm{tr} {\bf C}_S, \ I_2=\textrm{tr} ({\bf M} {\bf C}_S), \ I_3=\textrm{tr} ({\bf N} {\bf C}_S), \ I_4={\bf p} \cdot {\bf a}_1, I_5={\bf p} \cdot {\bf a}_2 \nonumber\\
&&I_6=\textrm{tr} {\bf b}_0, \ I_7=\textrm{tr} ({\bf M} {\bf b}_0), \ I_8=\textrm{tr} ({\bf N} {\bf b}_0).
\end{eqnarray}
Thus, for the energy we obtain
\begin{equation}
W=\tilde{W}(I_i), \ i=1,2,...,8. 
\end{equation}

By considering that the energy is linear with respect to the invariants we find the surface stresses
\begin{eqnarray}
{\bf S}_S=\frac{\partial W}{\partial {\bf C}_S}&&=\frac{\partial \tilde{W}}{\partial I_1} {\bf I}+\frac{\partial \tilde{W}}{\partial I_2} {\bf M}+\frac{\partial \tilde{W}}{\partial I_3} {\bf N} \nonumber\\
&&=\alpha {\bf I}+\beta {\bf M}+\gamma {\bf N}.
\end{eqnarray}
For the surface couple stress tensor we have
\begin{eqnarray}
{\bf m}_S=\frac{\partial W}{\partial {\bf b}_0}&&=\frac{\partial \tilde{W}}{\partial I_6} {\bf I}+\frac{\partial \tilde{W}}{\partial I_7} {\bf M}+\frac{\partial \tilde{W}}{\partial I_8} {\bf N}\nonumber\\
&&=\delta {\bf I}+\epsilon {\bf M}+\zeta {\bf N}.
\end{eqnarray}
For the term related with the shift vector we evaluate
\begin{eqnarray}
\frac{\partial W}{\partial {\bf p}}&&=\frac{\partial \tilde{W}}{\partial I_4} {\bf a}_1+\frac{\partial \tilde{W}}{\partial I_5} {\bf a}_2\nonumber\\
&&=\eta {\bf a}_1+\theta {\bf a}_1
\end{eqnarray}

Terms $I_1, I_6$ describe pure stretching and bending, respectively, effects. The effect of anisotropy to stretching and bending is described by terms $I_2, I_3$ and $I_7, I_8$, respectively. The effect of anisotropy to the shift vector is introduced throught terms $I_4, I_5$. All in all, the second breaking of symmetry requires specification of the 8 material parameters: $\alpha, \beta, \gamma,\delta, \epsilon, \zeta, \eta, \theta$. 

\section{Conclusion and future directions}

We present a nonlinear elastic constitutive framework for the modeling of 2D crystals of current interest such as graphene, hexagonal BN, MoS$_2$, WSe$_2$, and NbSe$_2$. We use the theory of monoatomic and diatomic 2-nets to find their arithmetic symmetries. Confined to weak transformation neighbourhoods and using the Cauchy-Born rule we are able to work with geometric symmetries. For finding the geometric symmetry group we evaluate the eigenvalues of the arithmetic symmetries and find the corresponding generators among the crystallographic point groups. We then apply the theory of invariants for an energy function depending on the surface Cauchy-Green deformation tensor, the curvature tensor and the shift vector. We lay down the expression for the stress tensor, the couple stress tensor as well as a term related with the shift vector. This is done for all case where symmetry changes due to applied loads. 

Future directions of our line of research are at two levels. Firstly, one important aspect is to find the symmetry breaking and symmetry preserving deformations for such materials in line with the approach of \cite{Pitteri-Zanzotto2003}. Secondly, we currently work on utilizing non-convex energies that can capture the phenomena at the transition regime, i.e., when symmetry breaks,  in line with fundamental works in zirconia (\cite{Truskinovsky-Zanzotto2002}); this might be expanded to the nonlinear case.

\section{Acknowledgements}
D. Sfyris sincerely thanks G. Zanzotto (Padova, Italy) for some very important clarifications concerning differences between monoatomic and diatomic 2-nets. This research has been co-financed by the European Union (European Social Fund - ESF) and Greek national funds through the Operational Program "Education and Lifelong Learning" of the National Strategic Reference Framework (NSRF) - Research Funding Program: ERC-10 "Deformation, Yield and Failure of Graphene and Graphene-based Nanocomposites". The financial support of the European Research Council through the projects ERC AdG 2013 (‘‘Tailor Graphene’’) is greatfully acknowledged.

\vspace{0.1cm}

D. Sfyris\\
FORTH-ICE/HT, Patras, Greece\\
dsfyris@iceht.forth.gr\\

G.I. Sfyris\\
LMS, Paris, France\\

C. Galiotis\\
FORTH-ICE/HT, and Deartment of Chemical Engineering, Patras, Greece\\

\end{document}